\documentclass[fleqn,twoside]{elsart3}
\usepackage{graphicx,amssymb}
\begin{document}

\begin{frontmatter}

\title{Magnetic Tunnel Window's Imprint and Beyond}

\author[label1]{Juan J. Alonso}\ead{jjalonso@uma.es}, 
\author[label2]{Julio F. Fern\'andez}\ead{jff@pipe.unizar.es} 
\address[label1]{Departamento F\'{\i}sica Aplicada,
Universidad de M\'alaga, M\'alaga, 29071 (Spain)}
\address[label2]{ICMA, CSIC and
Universidad de Zaragoza, 50009-Zaragoza (Spain)}


\begin{abstract}
We report results from Monte Carlo simulations of systems of magnetic dipoles
that relax through quantum tunneling, much as Fe$_8$ crystals at very
low temperature.
For short times, a hole develops in suitably defined magnetic field
pseudo--distributions, which
matches the shape of the tunnel window (TW). Much later, ordinary
field distributions
$P(h)$ develop similar holes if thermal energies are not much larger
than the TW's
energy. Still later, below the long--range ordering (LRO)
temperature, $P(h)$ exhibits the
signature of LRO.
\vspace{1pc}
\end{abstract}

\end{frontmatter}

Magnetic relaxation in crystals of single--molecule magnets, such as
Fe$_8$, has recently become a
subject of great interest. Experimentally observed relaxation at 
temperatures $T$
that are below some $10^{-2}$ of
single--ion crystal anisotropy barrier energies $U$ is temperature
independent, and is duly
attributted to magnetic quantum tunneling (MQT) \cite{sanww}. As
explained in Ref. \cite{PS},
hyperfine interactions with nuclear spins open up tunneling windows
(TW's) of energy
$\varepsilon_w$ that are comparable to magnetic dipolar energies. Then,
\begin{equation}
\Gamma^\prime (\varepsilon_h) \simeq
\Gamma\;\eta (\varepsilon_h/\varepsilon_w),
\label{gamma}
\end{equation}
is the tunneling rate
for spins at very low temperature, where $\Gamma$ is some rate, $\eta
(x)\sim 1$ for $\mid x \mid \lesssim
1$, $\eta (x)\sim 0$ for $x\gtrsim 1$, and $2 \varepsilon_h$ is the
Zeeman energy change upon tunneling. From here on, times are
given in terms of
$\Gamma^{-1}$. The results reported here in Figs. 1, 2(a), and 3 are
for $\eta (x)=1$ for $\mid x\mid <1$
and $\eta (x)=0$ for $\mid x\mid >1$; in Fig. 2(b) we show results
for $\eta (x)=\exp (-\mid x\mid )$. We
shall refer to the former TW as {\it flat} and to the latter one as
{\it exponential}.
\begin{figure}[!h]
\includegraphics*[width=70mm]{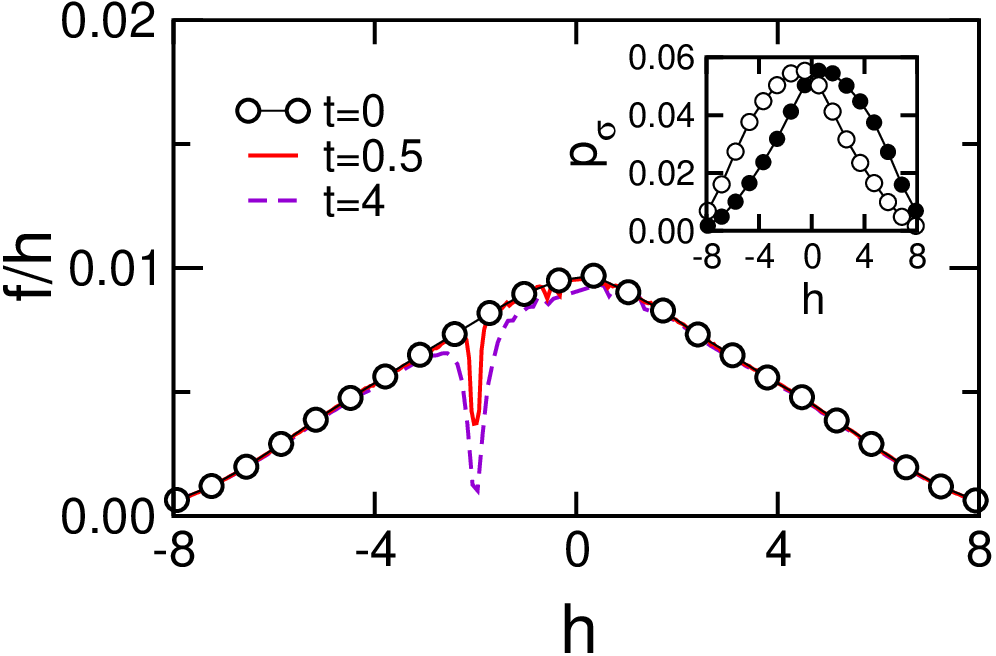}
\caption{\small $f/h$ versus $h$ for the shown times after quenching
and applying a field
$H=2$. All data points are for averages over $4\times 10^5$
independent runs for $\varepsilon_w=0.1$.
In the inset, $\bullet$ and $\circ$ are for $P_+$ and $P_-$.}
\end{figure}
An unusual magnetization $M$ relaxation of fully polarized Fe$_8$
like systems, $M_{t=0}-M_t\propto \sqrt
{t}$, was predicted in Ref. \cite{PS} and later observed \cite{exp}.
In an important piece of work,
Wernsdorfer et al. \cite{ww} found that
$M_t\propto \sqrt{t}$ if a small magnetic field $H$ is applied at
$t=0$, after quenching from a higher
temperature. This behavior has been explained recently \cite{nos}. In
addition, since $dM_t(H)/dt=-2 \Gamma \int dh
f(h,t)\eta(h+H)$ at all times, where
$f(h,t)=[P_+(h,t)-P_-(h,t)]$, and $P_+(h)$ [$P_-(h)$] is the density
distribution for fields
$h$ acting on up [down] spins, $f(h,t)$ can be
determined by a suitable application of magnetic
fields. [$f$ itself is not
normalized, and is therefore not a distribution.
In fact, it turns out that
$f(h)$ is proportional to the system's energy \cite{nos}]. A
``hole'' of ``intrinsic width'', was
thus observed to develop in time, and conjectured to correspond to
the width in $\eta (h)$, i.e., to the
width of the TW \cite{ww}.

We report Monte Carlo simulations of Ising systems of
interacting dipoles on SC lattices.
Initially, spins are allowed to flip readily at temperature $T_a\gtrsim
U/10\gtrsim T_{o}$ ($T_{o}\simeq 2.5$ is the LRO temperature) until the
energy per spin reaches some value $\varepsilon_a$. A field $H$ is
applied then and spins only flip at rate
$\Gamma^\prime$ [see Eq. (1)] thereafter, thus simulating
temperatures below some $10^{-2}U$.
We have previously found \cite {nos} that $M_t\propto \mid
\varepsilon_a\mid t$. Here, we report results
for the evolution of the field pseudo-distribution function $f(h)$ 
and for $p(H)\equiv
P_+(h)+P_-(h)$. Fields and energies are given in terms of nearest 
neighbor dipolar values.

How $P_+(h)$ and $P_-(h)$ split,
leading to a non-vanishing $f(h)$, after thermalization at high $T$
is illustrated in the inset of Fig. 1.
The hole that develops in $f(h)$ after quenching and applying field
$H$ is shown in Fig. 1. The shape of
the hole is shown in Figs. 2(a) and (b), for flat and exponential
TW's, respectively, for $H=2$.
Clearly, not only is the hole's width the same as the width of
$\eta$, as conjectured in Ref. \cite{ww},
but $f(h)/h$ matches $\eta (h)$ for the three shortest times shown in
Figs. 2(a) and (b). When $t\gtrsim
1$, the hole widens, as shown in Fig. 2(a) for $t=4$.

On the time scale on which the hole in $f/h$ develops, $P(H)$ remains
approximately independent of time.
On the other hand, a hole which resembles the TW grows at
much later times in $P(H)$ if
$T\lesssim \varepsilon_w$ \cite{nos2}. This is exhibited in Fig. 3.
Still later,
below the long--range ordering (LRO) temperature, $P(h)$ exhibits the
signature of LRO \cite{nos3} whether $T\lesssim \varepsilon_w$ or
not. If, on the other hand, the energy of the system is kept constant,
no LRO ensues, as expected.
This is illustrated in Fig. 3.
\begin{figure}[t]
\includegraphics*[width=7cm]{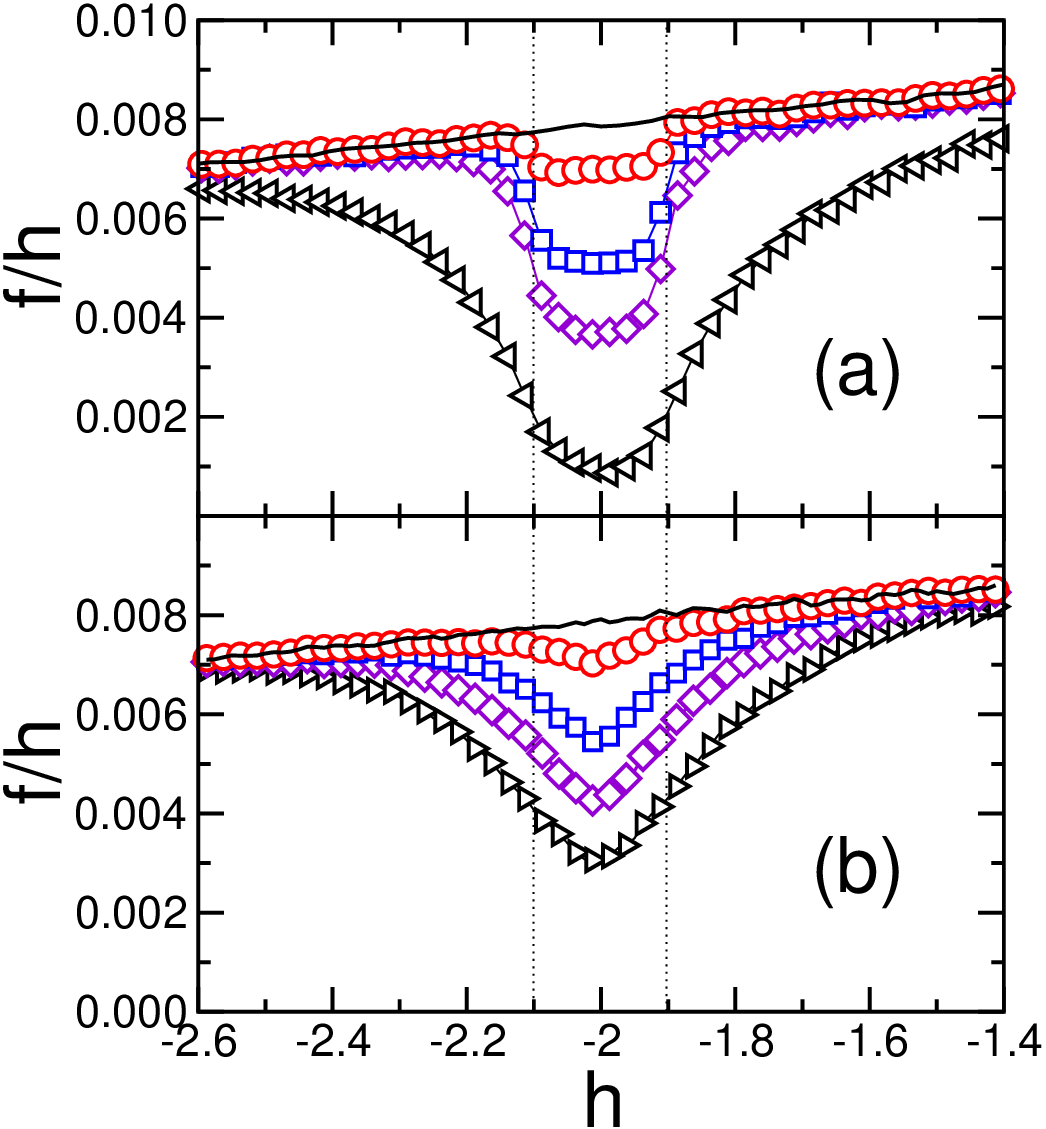}
\caption{\small (a) $f/h$ versus $h$ for short times with a flat TW
of $\varepsilon_w=0.1$,
$H=2$ and $\varepsilon_a=-0.51$  (b) Same as in (a) but for an
exponential, TW with
$\varepsilon_w=0.1$.  The solid line, $\circ$, $\square$, $\Diamond$
and $\triangleleft$ are for $t=0,
0.0625, 0.25, 0.5$ and $4$, respectively. }
\end{figure}

\begin{figure}[b]
\includegraphics*[width=7cm]{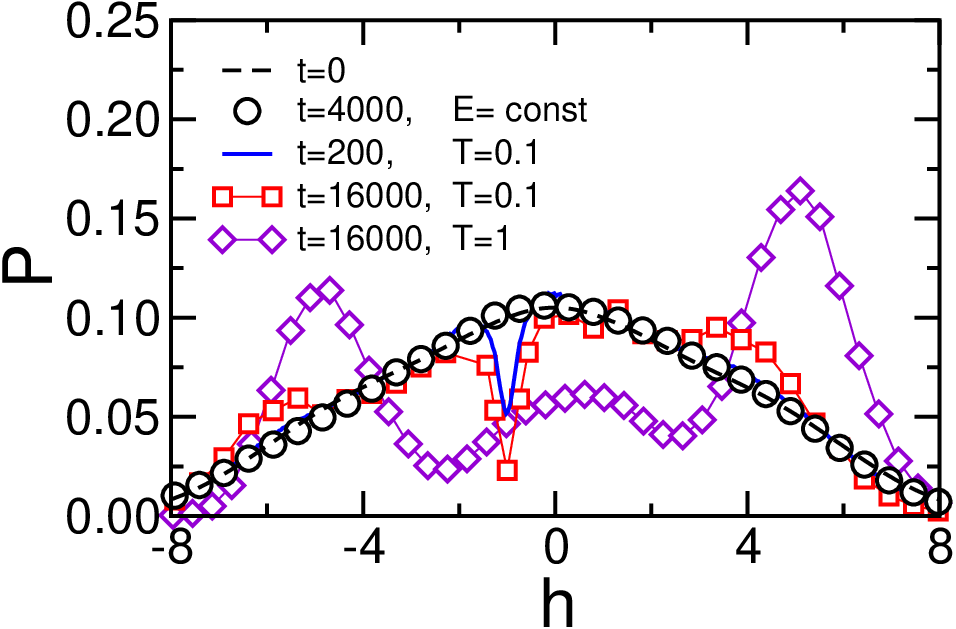}
\caption{\small The field distribution $P$ versus $h$ for the shown
times, and $T=0.1$ or constant energy.  In all cases
$\varepsilon_a=-0.58$, $\epsilon_w=0.2$ and $H=1$.}
\end{figure}

\end{document}